\documentclass[12pt]{article}
\voffset - 1 in
\usepackage{amsfonts}
\usepackage{amsmath}
\usepackage{amssymb,amsbsy}
\usepackage{bm}
\usepackage{makeidx}
\usepackage{latexsym}
\usepackage{graphicx}
\usepackage{epsfig}
\usepackage{subfigure}
\usepackage{dsfont}
\usepackage{mathcomp}
\usepackage{color}
\usepackage{mathrsfs}
\usepackage{epsf}
\newcommand{\beq}{\begin{eqnarray}}% can be used as {eqnarray}
\newcommand{\eeq}{\end{eqnarray}}
\newcommand{\be}{\begin{equation}}% can be used as {equation} 
\newcommand{\ee}{\end{equation}}

\begin{document}

\title{Mixed perturbations from an Inflaton and Two Curvatons - towards understanding non-gaussianity in more complicated curvaton scenarios}
\date{2nd August 2012}
\author{Maria Sueiro and Malcolm Fairbairn\\ Physics, King's College London, \\ Strand, London WC2R 2LS, UK}
\maketitle
\begin{abstract}
We analyse the situation where the primordial curvature perturbations are produced by the joint effects of an inflaton field and two curvaton fields.  We present general equations which allow the reader to obtain $f_{NL}$ for several different scenarios which differ in the order in which fields decay into radiation after inflation.  In order to investigate the physics of these equations we analyse some simplified situations where the fields are harmonic and both curvatons are frozen at the same expectation value during inflation.  We find quite complex behaviour -  for a given situation where the inflaton contributes a fixed amount to the total curvature perturbation there are situations where $f_{NL}$ is maximised if both curvatons share equally in contributing the rest and situations where $f_{NL}$ is maximised if only one of the two curvatons contributes the rest.  We are unable therefore to make any completely general extrapolations about the expected non-gaussianity from $N$ curvatons.  We find that as the curvaton contribution to the overall perturbation is gradually increased, $f_{NL}$ rises to a maximum before falling again and that a given $f_{NL}$ can correspond to many different parameter sets for the two curvatons.
\end{abstract}

\section{Introduction}

Inflation is the most successful theory we have to describe the early universe. According to inflation \cite{liddle}, at some early time the universe underwent an accelerated expansion, and this solves the horizon and flatness problem while also predicting an almost scale-invariant power spectrum in the cosmic microwave background, consistent with current observations. The most basic models of inflation (see, for example, \cite{lazarides}) involve a single, massive scalar field $\phi$ called the inflaton, which slowly rolls down its potential to source the necessary expansion. Once it reaches the minimum of the potential it oscillates around that point, reheating the Universe by decaying into standard model particles. The quantum fluctuations of the inflaton get stretched out with the expansion and become classical perturbations, causing inhomogeneities which upon horizon re-entry evolve to become the large-scale structure that we observe today \cite{liddlelyth}.

Since many theories that go beyond the standard model predict scalar fields, it is not unreasonable to assume that in early times, these fields had expectation values away from the minimum of their potential. Such fields, known as curvaton fields, can be present during inflation but only begin to move when the Hubble parameter drops down to their mass (as in \cite{lythwands} and \cite{lythungarelliwands}). They then proceed to oscillate around the minimum of their potential and eventually decay when the Hubble parameter is equal to their decay rate. The presence of these particles imprints a record of the curvaton perturbations as well as the inflaton perturbations upon the cosmic distribution of matter and therefore provides a different mechanism for the creation of the primordial curvature perturbation $\zeta$ and can alter the predictions of basic single-field inflation \cite{bartololiddle}, \cite{wandsmaliklythliddle}, \cite{byrneschoi}, \cite{wands}, \cite{dimopoulos}. There are a plethora of curvaton models one can imagine existing in particle physics and string theory and they typically include a large number of parameters. To constrain them, one needs to compare their predictions to current astrophysical observations, with the most popular discriminant being the non-gaussianity parameter, $f_{NL}$ (\cite{langlois}, \cite{bartolo},\cite{gordonlewis}, \cite{komatsu}).

The idea behind the curvaton scenario is to boost small perturbations and make them larger by virtue of the energy density in the curvaton field red-shifting more slowly than the energy density from the inflaton field.  This occurs once the inflaton has decayed into radiation but the curvaton is still oscillating (and thus red-shifting like matter).  There are therefore multiple scenarios depending upon the order in which the inflaton decays, the curvaton starts to move and the curvaton decays.  In this work we will attempt to study a number of these different situations, although we will not consider the case where the curvaton starts to roll during inflation, as this moves into the territory of multifield inflation.

In inflation models, the non-gaussianity parameter $f_{NL}$ is related to the curvature perturbation after the last decay. This parameter is a measure of deviations from a purely Gaussian distribution of the primordial curvature perturbation and the current $f_{NL}$ limits are given by WMAP 7 to be $-10<f_{NL}<74$ \cite{wmap7}. We analytically compute the expression for $\zeta$ at the last decay and $f_{NL}$ in models with one inflaton, $\phi$, and two curvatons, $\sigma$ and $\chi$, for various sequences of particle decays. In each case, the inflaton is the one responsible for producing the necessary e-folds and both the curvatons are frozen during the inflaton's slow-roll. Once inflation has finished, $\phi$  begins to oscillate around the minimum of its potential and then decays. Independently, some time after the end of inflation, when $H = m_{curvaton}$, the curvatons begin to oscillate and eventually decay.  All three particles contribute to the curvature perturbation.  We will evaluate the primordial curvature perturbation at the time of the last particle's decay, whether it is the inflaton or one of the curvatons.

\section{Perturbations and Non-Gaussianity with multiple fields}

In this section we will present our notation and derive some general expressions for the curvature perturbation $\zeta$ and the non-gaussianity parameter $f_{NL}$ which we will be able to use in the following sections.

\subsection{Curvature perturbation from the curvatons}

The simplest potential used in inflationary models is the quadratic, $V=\frac{1}{2}m_\phi^2\phi^2$ which we will use later on for our examples but in our equations we are allowing the curvaton potential to deviate from this case away from its minimum.

For each of the two curvatons, the oscillation amplitude will then be a function of the initial field expectation value at horizon exit, i.e. $\sigma=c(\sigma_{*})$ and $\chi=d(\chi_{*})$ respectively where the index $*$ means the expectation value at horizon exit.  This general case simplifies to the quadratic potential when $c=\sigma_{*} \Rightarrow c^\prime=1$ and $d=\chi_{*}\Rightarrow d^\prime=1$. The curvature perturbation up to second order for $\sigma$ is derived in \cite{sasakivaliviitawands} and it is
\beq
{\zeta_{\sigma}}_{(1)} &=& \frac{2}{3}\frac{\delta\sigma}{\bar{\sigma}},\\
{\zeta_{\sigma}}_{(2)} &=& -\frac{3}{2}\left(1 - \frac{c c^{\prime\prime}}{c^{\prime2}}\right){\zeta_{\sigma}}_{(1)}^2\label{socpc}
\eeq
the expressions for the other curvaton, $\chi$ will take a similar form.  
For each case that we study we will use $\zeta_\phi$, $\zeta_\sigma$ and $\zeta_\chi$ to define the isocurvature perturbation, $S_i\propto \zeta_i - \zeta_\phi $, $i=\sigma, \chi$, to describe the non-adiabatic part of the curvature perturbation of the curvatons. Next, we define the general curvature perturbation in terms of $\zeta_\phi$, $S_\sigma$ and $S_\chi$ to look at the non-gaussianity parameter $f_{NL}$.

\subsection{General expression for $f_{NL}$}

In this section, following a similar formalism as in \cite{assadullahi}, we derive the general expression for the $f_{NL}$ parameter in terms of the coefficients in the curvature perturbation produced in a model with one inflaton and two curvatons, taking into account contributions from all three particles.

Up to second order, the general form of the primordial curvature perturbation will be:
\begin{eqnarray}\label{gcp}
\zeta &=&{\rm A}{\zeta_{\phi}}_{(1)} + {\rm B} {S_{\sigma}}_{(1)} + \Gamma {S_{\chi}}_{(1)}\nonumber\\
&&+ \frac{1}{2}\Delta  {\zeta_{\phi}}_{(1)}^2 + \frac{1}{2}{\rm E}  {S_{\sigma}}_{(1)}^2 + \frac{1}{2}{\rm Z}  {S_{\chi}}_{(1)}^2\nonumber\\
&&+ \frac{1}{2}{\rm H}  {\zeta_{\phi}}_{(1)}{S_{\sigma}}_{(1)} + \frac{1}{2}\Theta  {\zeta_{\phi}}_{(1)}{S_{\chi}}_{(1)} + \frac{1}{2}{\rm I}  {S_{\sigma}}_{(1)}{S_{\chi}}_{(1)}
\end{eqnarray}
evaluated at the time of the last decay (the point at which all three fields have decayed into radiation). Since $f_{NL}$ is related to the primordial bi-spectrum, we compute the three point correlator of the curvature perturbation:
\begin{eqnarray}\label{cptsp}
\langle\zeta(\mathbf{k}_1)\zeta(\mathbf{k}_2)\zeta(\mathbf{k}_3)\rangle &=& \{[{\rm A}^2\Delta P_{\zeta_{\phi1}}(k_1)P_{\zeta_{\phi1}}(k_2) + {\rm B^2 E}P_{S_{\sigma1}}(k_1)P_{S_{\sigma1}}(k_2)\nonumber\\
&&+ \Gamma^2{\rm Z}P_{S_{\chi1}}(k_1)P_{S_{\chi1}}(k_2)+ \frac{1}{2}{\rm A B H}P_{\zeta_{\phi1}}(k_1)P_{S_{\sigma1}}(k_2)\nonumber\\
&&+ \frac{1}{2}{\rm A}\Gamma\Theta P_{\zeta_{\phi1}}(k_1)P_{S_{\chi1}}(k_2) + \frac{1}{2}{\rm B}\Gamma {\rm I}P_{S_{\sigma1}}(k_1)P_{S_{\chi1}}(k_2)]\nonumber\\
&& + \textrm{permutations}\} (2\pi)^3\delta^{(3)}(\mathbf{k_1}+\mathbf{k_2}+\mathbf{k_3})
\end{eqnarray}
where we have assumed that the adiabatic perturbation is uncorrelated with the isocurvature perturbations, so $\langle\zeta_{\phi_{(1)}}(\mathbf{k_1})S_{\sigma_{(1)}}(\mathbf{k_2})\rangle=\langle\zeta_{\phi_{(1)}}(\mathbf{k_1})S_{\chi_{(1)}}(\mathbf{k_2})\rangle=\langle S_{\sigma_{(1)}}(\mathbf{k_1})S_{\chi_{(1)}}(\mathbf{k_2})\rangle=0$. By looking at equation (\ref{gcp}), we can now see that the power spectrum for $\zeta$ at leading order is just given by
\begin{equation}\label{gps}
P_{\zeta} = {\rm A}^2 P_{\zeta_{\phi}} + {\rm B}^2 P_{S_{\sigma}} + \Gamma^2 P_{S_{\chi}}
\end{equation}
Where the power spectrum for each component is \cite{assadullahi},\cite{fonsecawands}, \cite{dimopoulos}:
\beq
P_{\zeta_{\phi}} & = & \frac{1}{2M_{Pl}^2\epsilon_*}\left(\frac{H_*}{2\pi}\right)^2\label{psphi}\\
P_{S_{\sigma}} & = & 4\left(\frac{c^\prime}{c}\right)^2\left(\frac{H_*}{2\pi}\right)^2\label{pssigma1}\\
P_{S_{\chi}} & = & 4\left(\frac{d^\prime}{d}\right)^2\left(\frac{H_*}{2\pi}\right)^2\label{pssigma2}
\eeq
here $\epsilon_*$ is the inflaton slow-roll parameter and $M_{Pl}$ is the reduced Planck mass, $M_{Pl}=(8\pi G)^{-1/2}$.  

Combining equations (\ref{gps}),(\ref{psphi}), (\ref{pssigma1}) and (\ref{pssigma2}) we can express all three individual components of the power spectrum $P_{\zeta_{\phi}}$, $P_{S_{\sigma}}$ and $P_{S_{\chi}}$ in terms of the total $P_{\zeta}$, and thus substitute them in the bi-spectrum equation (\ref{cptsp}). The non-linearity parameter is defined in the following way:
\beq
\langle\zeta(\mathbf{k}_1)\zeta(\mathbf{k}_2)\zeta(\mathbf{k}_3)\rangle&=&B(k_1, k_2, k_3)(2\pi)^3\delta^{(3)}(\mathbf{k_1}+\mathbf{k_2}+\mathbf{k_3})\nonumber\\
B(k_1, k_2, k_3)&=&\frac{6}{5}f_{NL}[P_\zeta(k_1)P_\zeta(k_2)+\textrm{permutations}]\nonumber.
\eeq
And so we are able to obtain the following equation for the $f_{NL}$ parameter:
\beq\label{gfnl}
f_{NL}&=&\frac{5}{6}\frac{1}{[{\rm A}^2c^2d^2 + 8{\rm B}^2M_{Pl}^2\epsilon_*c^{\prime2} d^2+ 8\Gamma^2M_{Pl}^2\epsilon_*c^2d^{\prime2}]^2}\times\{{\rm A}^2\Delta c^4d^4\nonumber\\
&& + 64{\rm B}^2{\rm E}M_{Pl}^4\epsilon_*^2c^{\prime4}d^4 + 64\Gamma^2{\rm Z}M_{Pl}^4\epsilon_*^2c^4d^{\prime4} + 4 {\rm ABH}M_{Pl}^2\epsilon_*c^2c^{\prime2}d^4\nonumber\\
&& + 4{\rm A}\Gamma\Theta M_{Pl}^2\epsilon_*c^4d^2d^{\prime2} + 32{\rm B}\Gamma{\rm I}M_{Pl}^4\epsilon_*^2c^2c^{\prime2}d^2d^{\prime2}\}
\eeq

This complicated looking expression is the general form of the non-linearity parameter when one inflaton and two curvatons are all contributing to the total curvature perturbation. The particular values of the parameters $A,B,\Gamma$ etc. depend upon the particular cosmic history. In the following sections we analytically compute the coefficients ${\rm A, B}, \Gamma,...$ for several different cases.  However before we do that we will check for consistency with previous results.

\subsubsection{First Consistency Check - $f_{NL}$ for curvatons without perturbations from inflation.\label{consis1}}

Before looking at the relatively complicated situation of mixed perturbations from an inflaton and two curvatons, we would like to check that our general equation (\ref{gfnl}) is able to re-create the results of others in particular limits.  The first such consistency check, i.e. the case where the perturbations come from two curvatons with no contribution from inflation, can be done immediately.  The other two consistency checks we will make (expression for a single curvaton and mixed perturbations from one inflaton and one curvaton) will have to wait until the parameters A,B,$\Gamma,\Delta$... etc have been defined in the next section.

If in our analysis above, we assume that the inflaton's contribution to the primordial curvature perturbation can be ignored, we have the case where the perturbations are coming from two curvatons alone.  In this case, ${\rm A}={\Delta}={\rm H}=\Theta=0$ and it is straightforward to see that the expression for $f_{NL}$ becomes
\be
f_{NL}=\frac{5}{6}\frac{{\rm B}^2{\rm E}+\frac{1}{2}\lambda^2{\rm B}\Gamma{\rm I}+\lambda^4\Gamma^2{\rm Z}}{({\rm B}^2+\lambda^2\Gamma^2)^2}
\ee where $\lambda=\sigma_*/\chi_*$. We have also assumed linear evolution of the two curvaton fields since horizon exit and so $c=\sigma_*$ and $d=\chi_*$. This is in agreement with the expression derived in \cite{assadullahi} for this particular case. 

\section{Mixed perturbations from an inflaton and two curvatons}

Now we move on to the main part of the analysis in this paper, namely the situation where the efolds of inflation required to solve the horizon problem are achieved through the stress energy of a single field, while the primordial perturbations are made up out of the individual curvature perturbations of all three fields.

We will calculate the primordial curvature perturbation after all the particles have decayed using the $\delta N$ formalism and the sudden decay approximation (for a different approach, using the ADM formalism see, \cite{lythmaliksasaki}). This means that we assume that as soon as the Hubble parameter $H$, becomes equal to the decay rate $\Gamma_i$ of a particle, the particle decays instantly and transfers all of its energy density (and curvature perturbation) into radiation. In our study we will follow the formalism of \cite{sasakivaliviitawands} and \cite{assadullahi}: we will obtain a chain of four equations to describe the evolution of our model starting from a time when just one particle has decayed through to the last particle's decay. Our goal is to find the first and second order components of the curvature perturbation at the last decay, expressed in terms of the initial adiabatic and isocurvature perturbations: $\zeta_\phi$, $S_\sigma$ and $S_\chi$. However, we generalise this method to include the linear terms coming from the inflaton, and up to second order terms from the two curvatons. Also, different orders of decay for the three particles will be considered, as well as situations where one of the curvatons is frozen until after the other two particles have decayed. Once we have the expression for $\zeta$, we can manipulate it in the general form of (\ref{gcp}) to find the coefficients we need to evaluate $f_{NL}$.

\subsection{Case 1: $t_{\sigma_{osc}} < t_{\chi_{osc}} < t_{\phi_{decay}} < t_{\sigma_{decay}} < t_{\chi_{decay}}$}

In this case, the two curvatons begin oscillating some time after the end of inflation, but decay after the inflaton has decayed. We are assuming that the energy density of each particle becomes radiation with the particle's decay. When a particle is oscillating, its equation of state is the one for pressure-less matter, $p=0$, while for a radiation component it is $p=\frac{1}{3}\rho$. So, after the inflaton decay but before the first curvaton decay (in other words, while the two curvatons are oscillating) the curvature perturbation and energy density of each species is
\beq
\zeta_{\phi} &=& \zeta_\alpha + \frac{1}{4}\ln(\frac{\rho_{\phi_{\alpha}}}{\bar{\rho}_{\phi_{\alpha}}})\Rightarrow \rho_{\phi_{\alpha}}=\bar{\rho}_{\phi_{\alpha}}\rm e^{4(\zeta_{\phi}-\zeta_\alpha)}\label{zetaphi}\\
\zeta_{\sigma} &=& \zeta_\alpha + \frac{1}{3}\ln(\frac{\rho_{\sigma_{\alpha}}}{\bar{\rho}_{\sigma_{\alpha}}})\Rightarrow \rho_{\sigma_{\alpha}}=\bar{\rho}_{\sigma_{\alpha}}\rm e^{3(\zeta_{\sigma}-\zeta_\alpha)}\label{zetasigma1}\\
\zeta_{\chi} &=& \zeta_\alpha + \frac{1}{3}\ln(\frac{\rho_{\chi_{\alpha}}}{\bar{\rho}_{\chi_{\alpha}}})\Rightarrow \rho_{\chi_{\alpha}}=\bar{\rho}_{\chi_{\alpha}}\rm e^{3(\zeta_{\chi}-\zeta_\alpha)}\label{zetasigma2}.
\eeq
The index $\alpha$ symbolises the time at which the first curvaton decays and we will use $\beta$ to denote the time at which the second curvaton decays, which will be the last decay of the three fields in all the scenarios analysed in this paper.  In the equations above, $\zeta_\alpha$ is the total curvature perturbation at the first curvaton decay and $\bar{\rho}_i$ is the average energy density of each field $i$. We can assume that at this point the total curvature perturbation is equal to the adiabatic perturbation and so equations (\ref{zetasigma1}) and (\ref{zetasigma2}) become
\beq
\rho_{\sigma_{\alpha}}=\bar{\rho}_{\sigma_{\alpha}}\rm e^{S_\sigma} , S_\sigma \equiv 3(\zeta_\sigma - \zeta_\phi)\label{Ssigma}\\
\rho_{\chi_{\alpha}}=\bar{\rho}_{\chi_{\alpha}}\rm e^{S_\chi} , S_\chi \equiv 3(\zeta_\chi-\zeta_\phi)\label{Schi}.
\eeq
We don't make use of the above definition for the two isocurvature perturbations immediately. We will use equations (\ref{Ssigma}) and (\ref{Schi}) once we obtain the curvature perturbation at the last decay to rewrite it into the form of (\ref{gcp}).

Since we are assuming non-linear evolution for the curvatons, $\bar{\rho}_\sigma=\frac{1}{2}m^2\bar{c}^2$, where by definition $\bar{c}=c(\bar{\sigma}_*)$, and the parameter $d$ is treated in an analogous way for the second curvaton $\chi$.
On a uniform density hypersurface just before the first curvaton decays, $\rho_{\phi_{\alpha}} + \rho_{\sigma_{\alpha}} + \rho_{\chi_{\alpha}}=\bar\rho_{{\textrm{total}}_{\alpha}}$. We can divide by $\bar\rho_{{\textrm{total}}_{\alpha}}$ and use equations (\ref{zetaphi}) to (\ref{zetasigma2}) to obtain
\be\label{meb1}
\Omega_{\gamma_0\alpha}e^{4(\zeta_{\phi}-\zeta_\alpha)} + \Omega_{\sigma\alpha} e^{3(\zeta_{\sigma}-\zeta_\alpha)} + \Omega_{\chi\alpha} e^{3(\zeta_{\chi}-\zeta_\alpha)} = 1
\ee
where $\Omega_i=\frac{\rho_{i_\alpha}}{\bar{\rho}_{\textrm{total}_\alpha}}$.  Here we replaced the index $\phi$ with $\gamma_0$ in the coefficient $\Omega$, since all of the inflaton's energy density is now the background radiation. We expand this equation to second order which yields three equations. The equation for zeroth order is
\be
\Omega_{\gamma_0\alpha} + \Omega_{\sigma\alpha} + \Omega_{\chi\alpha} = 1 , 
\ee
which is just an identity; the sum of all the energy densities is the total energy density. The first order terms give
\beq
{\zeta_\alpha}_{(1)} &=&  \frac{4\Omega_{\gamma_0\alpha}\zeta_{{\phi}_{(1)}}+3\Omega_{\sigma\alpha}\zeta_{{\sigma}_{(1)}}+3\Omega_{\chi\alpha}\zeta_{{\chi}_{(1)}}}
{4\Omega_{\gamma_0\alpha} + 3\Omega_{\sigma\alpha}+ 3\Omega_{\chi\alpha}}\nonumber\\
&\equiv&  f_{\gamma_{0\alpha}}{\zeta_{\phi}}_{(1)}  + f_{\sigma_\alpha}\zeta_{{\sigma}_{(1)}}+ f_{\chi_\alpha}\zeta_{{\chi}_{(1)}}\label{zeta11}
\eeq
where $f_{\gamma_{0\alpha}} + f_{\sigma_\alpha} + f_{\chi_\alpha} = 1$.
The second order part of equation (\ref{meb1}) is
\beq\label{zeta12}
{{\zeta_\alpha}_{(2)}}&=& 4f_{\gamma_{0\alpha}}({\zeta_{\phi}}_{(1)} - {\zeta_\alpha}_{(1)})^2  + 3f_{\sigma_\alpha}({\zeta_{\sigma}}_{(1)} - {\zeta_\alpha}_{(1)})^2 + 3f_{\chi_\alpha}({\zeta_{\chi}}_{(1)} - {\zeta_\alpha}_{(1)})^2\nonumber\\
&& + f_{\gamma_{0\alpha}}{\zeta_{\phi}}_{(2)}+ f_{\sigma_\alpha}{\zeta_{\sigma}}_{(2)} + f_{\chi_\alpha}{\zeta_{\chi}}_{(2)} .
\eeq

After the decay of the first curvaton, a similar analysis gives the equation
\be\label{mea1}
\Omega_{\gamma_1\alpha}e^{4(\zeta_{\gamma_1}-\zeta_\alpha)}+\Omega_{\chi\alpha} e^{3(\zeta_{\chi}-\zeta_\alpha)} = 1
\ee
where now the energy densities of both $\phi$ and $\sigma$ together constitute the radiation fluid, $\gamma_1$. Expanding this equation to second order gives the following for the zeroth order terms:
\be\label{eq5}
\Omega_{\gamma_1\alpha} + \Omega_{\chi\alpha} = 1 \Rightarrow \Omega_{\gamma_1\alpha} = 1- \Omega_{\chi\alpha}
\ee
which we use in the equation that we obtain from the first order part along with (\ref{zeta11}) to finally get the expression below for the incoming radiation perturbation at the decay of $\chi$:
\be\label{eq9}
{\zeta_{\gamma_1}}_{(1)} = R_1[1 - f_{\sigma_\alpha} - f_{\chi_\alpha}]{\zeta_{\phi}}_{(1)} + R_1 f_{\sigma_\alpha}{\zeta_{\sigma}}_{(1)} + [1 - R_1(1 - f_{\chi_\alpha})]{\zeta_{\chi}}_{(1)} ,
\ee
where $R_1=\frac{4 - \Omega_{\chi\alpha}}{4 - 4\Omega_{\chi\alpha}}$. We then look at the second order terms and, using equation (\ref{zeta12}), we eventually get:
\beq\label{eq19}
{\zeta_{\gamma_1}}_{(2)} &=& - 4({\zeta_{\gamma_1}}_{(1)} - {\zeta_\alpha}_{(1)})^2 + 3(1 - R_1)({\zeta_{\chi}}_{(1)} - {\zeta_\alpha}_{(1)})^2 + 4R_1f_{\gamma_{0\alpha}}({\zeta_{\phi}}_{(1)} - {\zeta_\alpha}_{(1)})^2\nonumber\\
&&  + 3R_1f_{\sigma_\alpha}({\zeta_{\sigma}}_{(1)} - {\zeta_\alpha}_{(1)})^2 + 3R_1f_{\chi_\alpha}({\zeta_{\chi}}_{(1)}- {\zeta_\alpha}_{(1)})^2 + R_1f_{\gamma_{0\alpha}}{\zeta_{\phi}}_{(2)}\nonumber\\
&& + R_1f_{\sigma_\alpha}{\zeta_{\sigma}}_{(2)} + [1 - R_1(1 - f_{\chi_\alpha})]{\zeta_{\chi}}_{(2)} .
\eeq

Now, considering the decay of the second curvaton $\chi$, symbolised by the index $\beta$, we obtain
\be\label{meb2}
\Omega_{\gamma_1\beta}e^{4(\zeta_{\gamma_1}-\zeta_\beta)}+\Omega_{\chi\beta} e^{3(\zeta_{\chi}-\zeta_\beta)} = 1,
\ee
valid just before $\chi$ decays and
\be\label{mea2}
\Omega_{\gamma_2\beta}e^{4(\zeta_{\gamma_2}-\zeta_\beta)}=1
\ee
valid right after the last decay. In the equation above, $\gamma_2$ is the radiation fluid after the last decay. 

Since all the particles have now decayed into radiation, $\zeta_\beta = {\zeta_\beta}_{(1)}+{\zeta_\beta}_{(2)}$ is the primordial curvature perturbation that we wish to evaluate in order to get $f_{NL}$. We want to find $\zeta_\beta$ in terms of the initial, known curvature perturbations of the individual fields. The coefficients are going to be functions of the particles' energy densities and of their ratios before the second decay and before the third decay.

We follow the same steps as in the study before and after the decay $\alpha$: we expand (\ref{meb2}) and (\ref{mea2}) to second order and with appropriate substitutions we find the first and second order part of the curvature perturbation at the time of the last decay in terms of ${\zeta_{\phi}}_{(1)}$, ${\zeta_{\sigma}}_{(1)}$ and ${\zeta_{\chi}}_{(1)}$.  
The first order component is:
\beq\label{fcpfo}
{\zeta_\beta}_{(1)} &=& R_1(1 - f_{\chi_\beta})(1 - f_{\sigma_\alpha} - f_{\chi_\alpha}){\zeta_{\phi}}_{(1)} + R_1f_{\sigma_\alpha}(1 - f_{\chi_\beta}){\zeta_{\sigma}}_{(1)}\nonumber\\
&& + [1 - R_1(1 - f_{\chi_\alpha})(1 - f_{\chi_\beta})]{\zeta_{\chi}}_{(1)}\\
&=& {\zeta_\phi}_{(1)} + \frac{1}{3}R_1 f_{\sigma_\alpha}(1-f_{\chi_\beta}){S_\sigma}_{(1)}+\frac{1}{3}[1-R_1(1-f_{\chi_\alpha})(1-f_{\chi_\beta})]{S_\chi}_{(1)}
\eeq
where to obtain the last line we used (\ref{Ssigma}) and (\ref{Schi}). We can already compare with (\ref{gcp}) to read off the first three coefficients $\rm{A}$, ${\rm B}$ and $\Gamma$:
\beq
{\rm A} &=& 1\\
{\rm B} &=& \frac{1}{3}R_1f_{\sigma_\alpha}(1 - f_{\chi_\beta})\\
\Gamma &=&\frac{1}{3}[1 - R_1(1 - f_{\chi_\alpha})(1 - f_{\chi_\beta})]
\eeq
The equation for the second order part is:
\beq\label{fcpso}
{\zeta_\beta}_{(2)}&=& 4(1 - f_{\chi_\beta})({\zeta_{\gamma_1}}_{(1)} - {\zeta_\beta}_{(1)})^2 + 3f_{\chi_\beta}({\zeta_{\chi}}_{(1)} - {\zeta_\beta}_{(1)})^2\nonumber\\
&&  + (1 - f_{\chi_\beta}){\zeta_{\gamma_1}}_{(2)} + f_{\chi_\beta}{\zeta_{\chi}}_{(2)}.
\eeq
We won't need any of the results coming from the expansion of (\ref{mea2}) but, for completeness, we included the calculation in the Appendix.
Having obtained equations (\ref{fcpfo}) and (\ref{fcpso}), we have the expression for the curvature perturbation just after the last decay,  which is the one we do need:
\beq\label{cfcp}
\zeta_\beta &=&{\zeta_\beta}_{(1)}+{\zeta_\beta}_{(2)}\nonumber\\
&=& R_1(1 - f_{\chi_\beta})(1 - f_{\sigma_\alpha} - f_{\chi_\alpha}){\zeta_{\phi}}_{(1)} + R_1f_{\sigma_\alpha}(1 - f_{\chi_\beta}){\zeta_{\sigma}}_{(1)}\nonumber\\
&& + [1 - R_1(1 - f_{\chi_\alpha})(1 - f_{\chi_\beta})]{\zeta_{\chi}}_{(1)} + 4(1 - f_{\chi_\beta})({\zeta_{\gamma_1}}_{(1)} - {\zeta_\beta}_{(1)})^2\nonumber\\
&& + 3f_{\chi_\beta}({\zeta_{\chi}}_{(1)} - {\zeta_\beta}_{(1)})^2 + (1 - f_{\chi_\beta}){\zeta_{\gamma_1}}_{(2)} + f_{\chi_\beta}{\zeta_{\chi}}_{(2)}.
\eeq
To be able to read off the necessary coefficients using (\ref{gcp}), we just need to write (\ref{cfcp}) in terms of the quantities ${\zeta_{\phi}}_{(1)}$, ${S_{\sigma}}_{(1)}$ and ${S_{\chi}}_{(1)}$. To achieve this we make use of equations (\ref{Ssigma}), (\ref{Schi}), (\ref{eq9}), (\ref{eq19}) and (\ref{fcpfo}). We can ignore the second order term of the inflaton's curvature perturbation ${\zeta_{\phi}}_{(2)}$, as in \cite{fonsecawands}, but we will keep the linear term so $\zeta_\phi=\zeta_{\phi_{(1)}}$. As for the curvatons, we consider first and second order terms, $\zeta_i=\zeta_{i_{(1)}}+\frac{1}{2}\zeta_{i_{(2)}}$, where $i=\sigma,\chi$ , keeping in mind equation (\ref{socpc}).

Below, we list the remaining 6 coefficients:
\beq
\Delta &=&\frac{3}{2}[-2+2R_1-R_1f_{\sigma_\alpha}(1-f_{\chi_\beta})-2R_1f_{\chi_\alpha}(1-f_{\chi2})+f_{\chi_\beta}-2R_1f_{\chi_\beta}\nonumber\\
&&+R_1f_{\sigma_\alpha}\frac{cc^{\prime \prime}}{{c^\prime}^2}+((1-R_1(1-f_{\chi_\alpha}))+f_{\chi_\beta})\frac{d d^{\prime \prime}}{{d^\prime}^2}]\\
{\rm E} &=& \frac{(1-f_{\chi_\beta})}{9}[4R_1^2f_{\sigma_\alpha}^2f_{\chi_\beta}^2 + 3R_1^2f_{\sigma_\alpha}f_{\chi_\beta}(1 - f_{\chi_\beta})- 4(1 - R_1)^2f_{\sigma_\alpha}^2\nonumber\\
&&+3(1 - R_1)f_{\sigma_\alpha}^2 + 4R_1f_{\sigma_\alpha}^2f_{\gamma_{0\alpha}}+3R_1f_{\sigma_\alpha}(1 - f_{\sigma_\alpha})^2\nonumber\\
&& + 3R_1f_{\sigma_\alpha}^2f_{\chi_\alpha} - \frac{3}{2}R_1f_{\sigma_\alpha}(1 - \frac{cc^{\prime\prime}}{c^{\prime2}})]\\
{\rm Z} &=&\frac{(1-f_{\chi_\beta})}{9}[4R_1^2f_{\sigma_\alpha}^2f_{\chi_\beta}^2 + 3R_1^2f_{\sigma_\alpha}^2(1 - f_{\chi_\beta})f_{\chi_\beta} - 4(1 - R_1)^2(1-f_{\sigma_\alpha})^2\nonumber\\
&&+3(1 - R_1)(1 - f_{\chi_\alpha})^2 + 4R_1f_{\gamma_{0\alpha}}f_{\chi_\alpha}^2f_{\chi_\beta} + 3R_1f_{\sigma_\alpha}f_{\chi_\alpha}^2\nonumber\\
&&+\frac{3}{2}R_1f_{\chi_\alpha}(1-f_{\chi_\alpha})^2(1 - \frac{dd^{\prime\prime}}{d^{\prime2}})] - \frac{3}{2}(1 - \frac{dd^{\prime\prime}}{d^{\prime2}})\\
{\rm H} &=& -R_1f_{\sigma_\alpha}(1-f_{\chi_\beta})+ R_1f_{\sigma_\alpha}\frac{cc^{\prime \prime}}{{c^\prime}^2}\\
\Theta &=& -2+2R_1-2R_1f_{\chi_\alpha}(1-f_{\chi_\beta})+f_{\chi_\beta}-2R_1f_{\chi_\beta}+((1-R_1(1-f_{\chi_\alpha}))+f_{\chi_\beta})\frac{d d^{\prime \prime}}{{d^\prime}^2}\\
{\rm I} &=& \frac{(1-f_{\chi_\beta})}{9}[-8R_1^2f_{\sigma_\alpha}(1-f_{\chi_\alpha})-6R_1^2f_{\sigma_\alpha}(1-f_{\chi_\alpha})f_{\chi_\beta}(1-f_{\chi_\beta})\nonumber\\
&&+8(1-R_1)^2f_{\sigma_\alpha}(1-f_{\sigma_\alpha})-6(1-R_1)f_{\sigma_\alpha}(1-f_{\chi_\alpha})\nonumber\\
&&+8R_1f_{\gamma_{0\alpha}}f_{\sigma_\alpha}f_{\chi_\alpha}-6R_1f_{\sigma_\alpha}f_{\chi_\alpha}(1-f_{\sigma_\alpha})\nonumber\\
&&-6R_1f_{\sigma_\alpha}f_{\chi_\alpha}(1-f_{\chi_\alpha})]
\eeq
We now have everything we need to evaluate the non-linearity parameter for the given configuration, making use of equation (\ref{gfnl}). Apart from the nine coefficients that we have evaluated above, we need to use the Planck mass $M_{Pl}=2.4 \times 10^{18}GeV$, the first slow-roll parameter $\epsilon_*=\frac{1}{2}M_{Pl}^2(\frac{V^\prime}{V})^2$ as well as $c$ and $d$. In the case of a quadratic potential for both the curvatons, it is simply $c=\sigma$ and $d=\chi$.

\subsubsection{Second Consistency Check - $f_{NL}$ for a single curvaton.}

Having derived these parameters for this case, we are now in a position to be able to make two more consistency checks which compare our results to previous work.  We can assume that the second curvaton $\chi$ is not present and that the contribution to the perturbation from the radiation field (i.e. the contribution from the inflaton) is equal to zero $\zeta_\gamma=0$.  We have ${\rm A}=\Gamma=\Delta={\rm Z}={\rm H}=\Theta={\rm I}=0$, and (\ref{gfnl}) becomes
\be
f_{NL}=\frac{5}{4f_{\sigma_\alpha}}\left(1+\frac{cc^{\prime\prime}}{c^{\prime2}}\right)-\frac{5}{3}-\frac{5}{6}f_{\sigma_\alpha}
\ee  
which is the same as the single curvaton expression found in \cite{sasakivaliviitawands} (and in \cite{lythrodriguez} for $cc^{\prime\prime}/c^{\prime2}=0$).  

Note that the fractional contributions to the density of the radiation coming from the decay (e.g. $f_{\sigma_\alpha}$) depend in the simple harmonic case upon the mass, the decay rate and the expectation value during inflation of the curvaton field.  In both of the situations we looked at above (here and in section \ref{consis1}), the non-gaussianity does not therefore depend upon the expectation value of the curvaton field explicitly, although in the case of two curvatons, it does depend upon the ratio of the expectation values.  

\subsubsection{Third Consistency Check - $f_{NL}$ for mixed perturbations from one inflaton and one curvaton}

\begin{figure}[!htbp]
\centering
\begin{tabular}{cc}
\includegraphics[width=70mm, height=70mm]{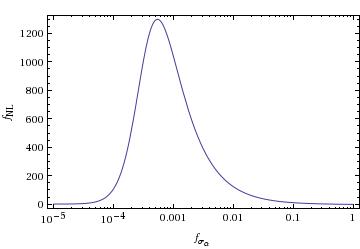}& 
\includegraphics[width=70mm, height=70mm]{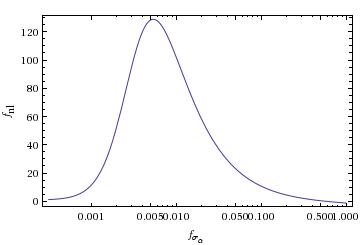}\\
\end{tabular}
\caption{The non-gaussianity $f_{NL}$ for $f_{\sigma_\alpha}\in[0,1]$, for $\sigma_*=10^{15}GeV$ and $\sigma_*=10^{16}GeV$ in the 1 inflaton-1 curvaton model.}
\label{fig:fnlphisigma1}
\end{figure}

\begin{figure}[!htbp]
\centering
\begin{tabular}{cc}
\includegraphics[width=70mm, height=70mm]{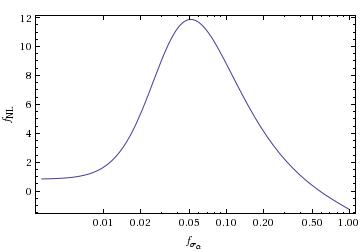}&
\includegraphics[width=70mm, height=70mm]{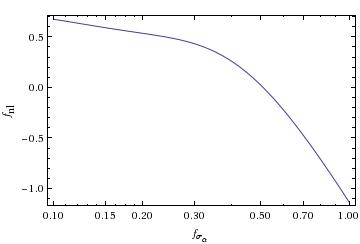}\\
\end{tabular}
\caption{The non-gaussianity $f_{NL}$ for $f_{\sigma_\alpha}\in[0,1]$, for $\sigma_*=10^{16}GeV$ on the left and $\sigma_*=10^{17}GeV$ on the right in the 1 inflaton-1 curvaton model.}
\label{fig:fnlphisigma2}
\end{figure}

If we assume that only the inflaton $\phi$ and a single curvaton $\sigma$ are present, we immediately see that $\Gamma={\rm Z}=\Theta={\rm I}=0$ and that the rest of the coefficients take much simplified forms. In the end, our expression for $f_{NL}$ is reduced to:
\beq
f_{NL}&=&\frac{5}{6[9\sigma_*^2 + 8M_{Pl}^2\epsilon_* f_{\sigma_\alpha}^2]^2}\times\{9\sigma_*^4 - \frac{32}{9}M_{Pl}^4\epsilon_*^2f_{\sigma_\alpha}^3(-3+4f_{\sigma_\alpha}+2f_{\sigma_\alpha}^2)\\
&& + 4M_{Pl}^2\epsilon_*\sigma_*^2f_{\sigma_\alpha}[2f_{\sigma_\alpha}(-3+2f_{\sigma_\alpha}+f_{\sigma_\alpha}^2)-f_{\sigma_\alpha}(-3+4f_{\sigma_\alpha}+2f_{\sigma_\alpha}^2)]\}
\eeq
In the situation where there are two curvatons and no inflaton, the non-gaussianity depends upon the ratio of the two expectation values of the curvatons.  In this case of one inflaton and one curvaton, the expression for $f_{NL}$ depends on the curvaton expectation value; we study it for  $\sigma_*=10^{14}GeV$,$\sigma_*=10^{15}GeV$, $\sigma_*=10^{16}GeV$ and $\sigma_*=10^{17}GeV$ in plots ~\ref{fig:fnlphisigma1} and ~\ref{fig:fnlphisigma2}.

For the first three cases we notice that the non-gaussianity increases as $f_{\sigma_\alpha}$ becomes smaller, but not indefinitely. Instead, $f_{NL}$ has a maximum value of ${f_{NL}}_{max}=1298$, ${f_{NL}}_{max}=128.8$ and ${f_{NL}}_{max}=11.9$ respectively.  So if we were to observe a non-gaussianity of $f_{NL}=50$ which we were to try to interpret in terms of mixed perturbations from a single inflaton and a single curvaton both with harmonic potentials, there would be two combinations of mass/decay time which would correspond to the same $f_{NL}$. For $\sigma_*=10^{17}GeV$ we see in Figure ~\ref{fig:fnlphisigma2} that the non-gaussianity starts from values around -1 and as $f_{\sigma_\alpha}$ becomes smaller it increases and reaches a maximum of ${f_{NL}}_{max}=0.83$ around which it then stays constant.

This is the same model as the one studied recently in \cite{fonsecawands}; we see the same capacity of the model to produce large values of $f_{NL}$ and the same increase of the non-gaussianity parameter when the curvaton expectation value becomes smaller. We also find that $f_{NL}$ has a maximum possible value and it in fact returns to the region allowed by observations after $f_{\sigma_\alpha}$ is smaller than $2 \times 10^{-3}$ for $\sigma=10^{15}GeV$. Our Figures ~\ref{fig:fnlphisigma1} and ~\ref{fig:fnlphisigma2} are in agreement with their Fig. 2.

\subsubsection{Case 1 with simultaneous decay}

We would like to put some numbers into our equations to see what they are telling us.  In this section, we will evaluate the $f_{NL}$ under the simplifying assumption that the second and third decay of case 1 happen simultaneously.  Furthermore, we assume that both curvatons have a perfectly quadratic potential so that $cc^{\prime\prime}/c^{\prime 2} = 0$ and $dd^{\prime\prime}/d^{\prime 2} = 0$.  To completely simplify the situation we assume that the two curvatons have equal expectation values at horizon crossing, $\sigma_*=\chi_*$. In our plots for $f_{NL}$ we use the reduced Planck mass $M_{Pl}=2.4\times10^{18}GeV$ and the slow roll parameter $\epsilon_*\approx0.02$. 

\begin{figure}[!t]
\centering
\begin{tabular}{cc}
\includegraphics[width=70mm, height=70mm]{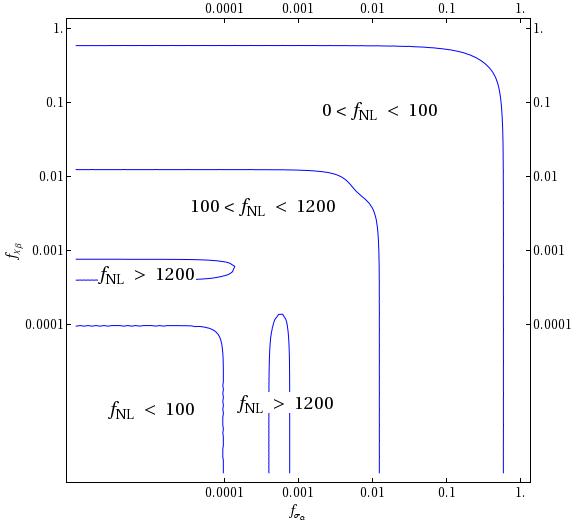}& 
\includegraphics[width=70mm, height=70mm]{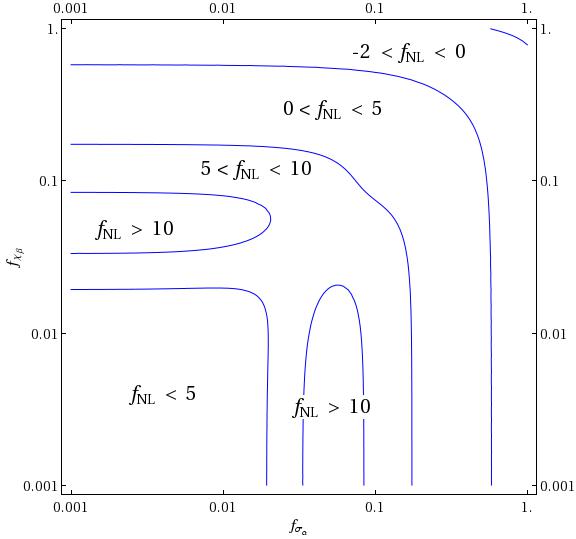}\\
\end{tabular}
\caption{Contours for $f_{NL}(f_{\sigma_\alpha}, f_{\chi_\beta})$ as predicted in a simplified version of Case 1 for $\sigma_*=10^{14}GeV$, $f_{\sigma_\alpha}\in[0,1]$ and $f_{\chi_\beta}\in[0,1]$ on the left. In the second plot, $\sigma_*=10^{16}GeV$ while the range is  $f_{\sigma_\alpha}\in[10^{-3},1]$ and $f_{\chi_\beta}\in[10^{-3},1]$.}
\label{fig:fnl1}
\end{figure}

After the inflaton $\phi$ decays, the two curvatons decay simultaneously, $\Omega_{\chi\alpha}=\Omega_{\chi\beta}$. This means that $t_\alpha=t_\beta$, but to remain consistent with our notation we will express the curvature perturbation coefficients in terms of $f_{\sigma_\alpha}$ and $f_{\chi_\beta}$. In reference \cite{assadullahi}, we find that for simultaneous decay of the two curvatons the following equation holds:
\be
f_{\chi_\alpha}=(1+\frac{f_{\sigma_\alpha}}{3})f_{\chi_\beta} .
\ee
which immediately significantly simplifies the nine coefficients (A,B,$\Gamma$...) in the expression for $\zeta$ (\ref{gcp}). The expression above is a direct result of the fact that we can consider the value of $f_{\chi_\alpha}$ as constrained by the given values of $f_{\sigma_\alpha}$ and $f_{\chi_\beta}$. Since $f_{\gamma_0\alpha}+f_{\sigma_\alpha}+f_{\chi_\alpha}=1$ we can write the inequality $f_{\chi_\alpha}\leq1-f_{\sigma_\alpha}$ and this eventually leads to $f_{\chi_\alpha}\leq(1+\frac{f_{\sigma_\alpha}}{3})f_{\chi_\beta}$ which is valid for all configurations. 
We use the simplified coefficients from equations (\ref{c11}) to (\ref{c19}) and equation (\ref{gfnl}) to obtain Figure ~\ref{fig:fnl1} for $f_{\sigma_\alpha}\in[0,1]$ and $f_{\chi_\beta}\in[0,1]$.

For $\sigma_*=10^{14}GeV$, we find that the maximum value for the non-linearity parameter is ${f_{NL}}_{max}=1298$. This value of $f_{NL}$ is obtained when $f_{\chi_\beta}$ goes to zero (or alternatively $f_{\sigma_\alpha} \rightarrow 0$ since this case is symmetric), while the contribution of the other curvaton is $\sim 5 \times 10^{-4}$. This is precisely the same maximum value of $f_{NL}$ as the one in Figure ~\ref{fig:fnlphisigma1} where we studied a model with one inflaton and one curvaton, as would be expected.It's the same in the case of $\sigma_*=10^{16}GeV$; we find that the maximum value of the non-gaussianity is ${f_{NL}}_{max} = 11.9$ and is identical to the result we got in the 1 inflaton 1 curvaton case. This maximum value is not observed in the non-gaussianity produced in the two curvaton scenario as studied in \cite{assadullahi}. The only difference between our Case 1 and the model in \cite{assadullahi} is the inclusion of the inflaton's contribution to the perturbations in our work, which creates an upper boundary for $f_{NL}$. 

\begin{figure}[!htbp]
\centering
\includegraphics[width=70mm, height=70mm]{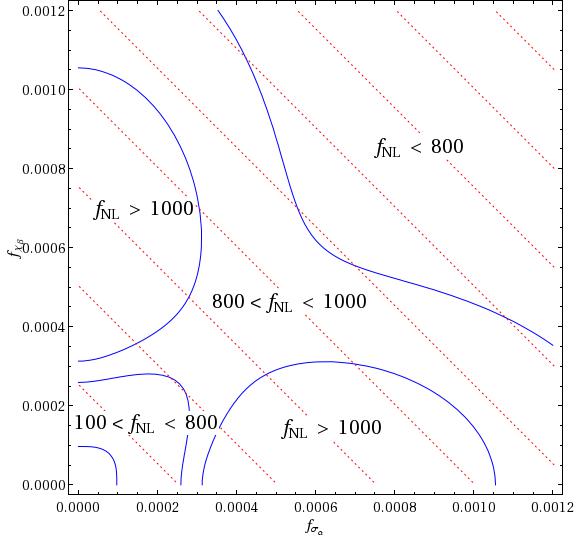}
\caption{The area of large $f_{NL}$ in case 1 for $\sigma_*=10^{14}GeV$. The red dashed lines correspond to $f_{\sigma_\alpha}+f_{\chi_\beta}=\textrm{constant}$. }
\label{fig:fnl1const}
\end{figure}

It is evident in all our plots that there are many different combinations of $f_{\sigma_\alpha}$ and $f_{\chi_\beta}$ that can give the same prediction for non-gaussianity. On the other hand, we want to emphasise with Figure ~\ref{fig:fnl1const}, that even when we keep the total contribution of the curvatons to the perturbation constant (red dashed lines), the $f_{NL}$ can still change significantly.

\subsection{Case 2: $t_{\sigma_{osc}} < t_{\chi_{osc}} < t_{\sigma_{decay}} < t_{\phi_{decay}} < t_{\chi_{decay}}$}

For the second case, we assume that one of the curvatons decays early, before the inflaton. Following the same procedure as in the first case we write down an equation valid just before the inflaton decays (decay 1):
\be\label{meb1c2}
\Omega_{\gamma_0\alpha}e^{4(\zeta_{\sigma}-\zeta_\alpha)} + \Omega_{\phi_1} e^{3(\zeta_{\phi}-\zeta_\alpha)} + \Omega_{\chi\alpha} e^{3(\zeta_{\chi}-\zeta_\alpha)} = 1 .
\ee
Here, the energy density of the curvaton $\sigma$, which has decayed, forms the background radiation $\gamma_0$, while the other two particles are still oscillating so their energy density evolves like matter.  This means that in what follows in this section, $\Omega_{\gamma_0\alpha}=\Omega_{\sigma\alpha}$ and $f_{\gamma_{0\alpha}}=f_{\sigma_\alpha} = 1-f_{\phi_\alpha} - f_{\chi_\alpha}$.
Once the inflaton decays as well, we have
\be\label{mea1c2}
\Omega_{\gamma_1\alpha}e^{4(\zeta_{\gamma_1}-\zeta_\alpha)}+\Omega_{\chi\alpha} e^{3(\zeta_{\chi}-\zeta_\alpha)} = 1
\ee
where the energy densities of $\phi$ and $\sigma$ have been converted to radiation, and we see that the system is now the same as in the first case: the inflaton and one curvaton have decayed into radiation, while the second curvaton oscillates. This means that  in this case the equations valid before and after the last decay are simply (\ref{meb2}) and (\ref{mea2}) respectively.
To obtain the coefficients present in the non-gaussianity expression (\ref{gfnl}), we need the first and second order components of the curvature perturbation at the last decay. Just like we did in the previous case we expand the equations above to second order, to eventually get
\beq
\zeta_\beta &=& R_1f_{\phi_1}(1-f_{\chi_\beta}){\zeta_{\phi}}_{(1)} + R_1f_{\sigma_\alpha}(1-f_{\chi_\beta}){\zeta_{\sigma}}_{(1)}\nonumber\\
&&+[1-R_1(1-f_{\chi_\alpha})(1-f_{\chi_\beta})]{\zeta_{\chi}}_{(1)} + 4(1-f_{\chi_\beta})({\zeta_{\gamma_1}}_{(1)}-{\zeta_\beta}_{(1)})^2\nonumber\\
&& + 3f_{\chi_\beta}({\zeta_{\chi}}_{(1)}-{\zeta_\beta}_{(1)})^2 + (1-f_{\chi_\beta}){\zeta_{\gamma_1}}_{(2)}+f_{\chi_\beta}{\zeta_{\chi}}_{(2)} ,
\eeq
which we can rewrite solely in terms of ${\zeta_{\phi_1}}_{(1)}$, ${S_{\sigma}}_{(1)}$ and ${S_{\chi}}_{(1)}$ to read off the curvature perturbation coefficients. The list can be found in equations (\ref{c21})-(\ref{c29}) in the Appendix.

\subsubsection{Case 2 with simultaneous decay}
Again, we would like to evaluate our equations to find out what new information they are giving us.

\begin{figure}[!htbp]
\centering
\begin{tabular}{cc}
\includegraphics[width=70mm, height=70mm]{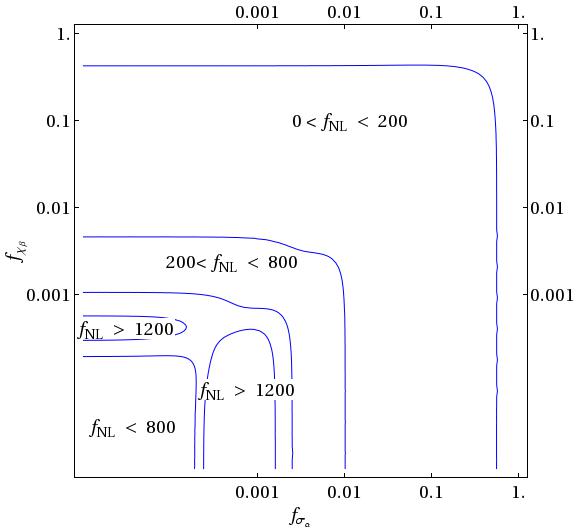}& 
\includegraphics[width=70mm, height=70mm]{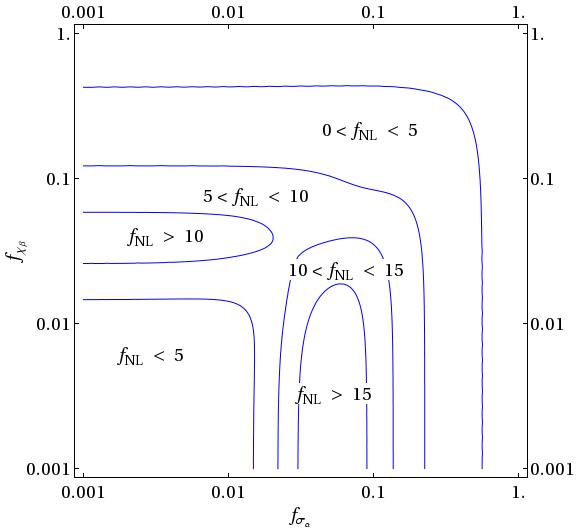}\\
\end{tabular}
\caption{$f_{NL}$ in Case 2 when the second and third decay happen simultaneously, in terms of $f_{\sigma_\alpha}$ and $f_{\chi_\beta}$. The plot on the left is done for $\sigma_*=10^{14}GeV$ with $f_{\sigma_\alpha}\in[0,1]$ and $f_{\chi_\beta}\in[0,1]$, while the plot on the right for $\sigma_*=10^{16}GeV$ with $f_{\sigma_\alpha}\in[10^{-3},1]$ and $f_{\chi_\beta}\in[10^{-3},1]$.}
\label{fig:fnl21}
\end{figure}

In this case we assume that after curvaton $\sigma$ decays, the inflaton and the other curvaton decay at the same time.  The coefficients in the curvature perturbation expression written in terms of $f_{\sigma_\alpha}$ and $f_{\chi_\beta}$ can be obtained by placing the relation $f_{\gamma_{0\alpha}}=f_{\sigma_\alpha} = 1-f_{\phi_\alpha} - f_{\chi_\alpha}$ into equations (\ref{c21})-(\ref{c29}) to produce equations (\ref{c2sd1}) to (\ref{c2sd9}).  We then create contour plots of our expression for $f_{NL}$ in Figure ~\ref{fig:fnl21}. The maximum value for $f_{NL}$ is ${f_{NL}}_{max}=2162.5$ for $\sigma_*=10^{14}GeV$.  This is significantly larger than the maximum $f_{NL}$ value of Case 1, which shows that the order of the particles' decay can have a big effect on non-gaussianity.

 We see that the prediction for $f_{NL}$ is significantly smaller for larger $\sigma_*$. In the case where $\sigma_*=10^{16}GeV$ we find the maximum value ${f_{NL}}_{max}=19.2$.

\subsection{Case 3:  $t_{\sigma_{osc}} < t_{\chi_{osc}} < t_{\sigma_{decay}} < t_{\chi_{decay}}< t_{\phi_{decay}}$}
In the third scenario, we consider a late decaying inflaton, decaying after both the curvatons have oscillated and decayed. The decay of the curvaton $\sigma$ creates the radiation fluid present just before the decay of $\chi$:
\be\label{meb1c3}
\Omega_{\gamma_0\alpha}e^{4(\zeta_{\sigma}-\zeta_\alpha)} + \Omega_{\chi\alpha} e^{3(\zeta_{\chi}-\zeta_\alpha)} + \Omega_{\phi_1} e^{3(\zeta_{\phi}-\zeta_\alpha)}= 1 .
\ee
After the second curvaton decays as well, we have:
\be\label{mea1c3}
\Omega_{\gamma_1\alpha}e^{4(\zeta_{\gamma_1}-\zeta_\alpha)}+\Omega_{\phi_1} e^{3(\zeta_{\phi}-\zeta_\alpha)} = 1 .
\ee
We can also write the following two equations valid before and after the last decay:
\beq
\Omega_{\gamma_1\beta}e^{4(\zeta_{\gamma_1}-\zeta_\beta)}+\Omega_{\phi_1} e^{3(\zeta_{\phi}-\zeta_\beta)} &=& 1\label{meb2c3}\\
\Omega_{\gamma_2\beta}e^{4(\zeta_{\gamma_2} - \zeta_\beta)} &=& 1\label{mea2c3} .
\eeq
These are the four main equations for this case and expanding them to second order eventually results in the expression for the primordial curvature perturbation at the time of last decay
\beq
\zeta_\beta &=& [1 - R_1(1-f_{\phi_\alpha})(1-f_{\phi_\beta})]{\zeta_{\phi}}_{(1)} + R_1f_{\sigma_\alpha}(1 - f_{\phi_\beta}){\zeta_{\sigma}}_{(1)}\nonumber\\
&& + R_1f_{\chi_\alpha}(1-f_{\phi_\beta}){\zeta_{\chi}}_{(1)} +4f_{\phi_\beta}({\zeta_{\gamma_1}}_{(1)}-{\zeta_\beta}_{(1)})^2\nonumber\\
&& + 3(1-f_{\phi_\beta})({\zeta_{\phi}}_{(1)} - {\zeta_\beta}_{(1)})^2 + f_{\phi_\beta}{\zeta_{\gamma_1}}_{(2)} + (1-f_{\phi_\beta}){\zeta_{\phi}}_{(2)}
\eeq
We rewrite this to match the form of equation (\ref{gcp}). We list the resulting nine coefficients in the Appendix.

Case 3, in the simultaneous decay limit is equivalent to Case 2 for $b \leftrightarrow \phi$.

\section{Discussion}
In this work, we have studied the non-gaussianity predicted in models of inflation with 1 inflaton and 2 curvatons. We derived a general expression for $f_{NL}$ from the bi-spectrum of a general curvature perturbation $\zeta$. We saw that our general expression simplifies to reproduce other expressions for $f_{NL}$ in the literature under the relevant simplifications. 

We then analytically derived the necessary coefficients for specific scenarios in which we varied the sequence of the decays of the three particle species. For our main three cases, we then simplified our $f_{NL}$ expression under the assumption that the last two decays happen at the same time. 

For all of the calculations we performed in this paper, we assumed a harmonic potential for the inflaton of the form $m^2\phi^2$ and we assumed therefore a slow roll parameter 60 efolds before the end of inflation of $\epsilon\sim 0.02$. We made no assumption for the potential of the curvatons in our general calculations; for the simplified cases we used the quadratic form.

In the simultaneous decay limit we studied the features of the predicted $f_{NL}$ and found that the produced non-gaussianity is bounded. When the contribution of the curvaton to the final perturbation is of order unity, $f_{NL}$ is slightly negative but becomes more positive as $f_{\sigma_\alpha}$ and $f_{\chi_\beta}$ become smaller.  The maximum value of $f_{NL}$ is set in this situation by the expectation value $\sigma_*$ of the curvaton during inflation and is smaller when $\sigma_*$ is larger.  For expectation values smaller than around $10^{16}$ GeV $f_{NL}$ grows beyond the current observational constraints if the relative contribution of the curvaton has the right value. If the expectation value of the curvaton is $10^{17}$ GeV then we find that the maximum non-gaussianity is around $f_{NL}\sim 0.8$ which is not in the range that will be detectable by Planck.

We find that $f_{NL}$ can take very large values, depending upon the expectation value of the curvaton during inflation but that it does not grow indefinitely. After it reaches a maximum it decreases and returns to within the range allowed by current observations. 

A particular value of $f_{NL}$ therefore corresponds to two different sets of parameters, even in the situation where there is a single inflaton and a single curvaton and more information is required.

The predicted $f_{NL}$ in all of the three field models in this paper is quite a complicated function of the relative contributions of the two curvatons.  If, for example, we take the situation where 80\% of the total curvature perturbation comes from the inflaton and 20\% comes from the combined effect of the two curvatons, there are situations where having all of the curvaton contribution to the pertubation coming from one of the two curvaton fields will maximise $f_{NL}$, and there are situations where having 10\% of the perturbation coming from each curvaton will maximise $f_{NL}$, depending upon the expectation values of the curvaton fields.  There are therefore in general several combinations of field parameters that give the same non-gaussianity and, alternatively, even when these parameters change in a way that leaves the total curvaton contribution to $\zeta$ constant there can be big changes in the value of $f_{NL}$.

Of course it would be interesting therefore to investigate the trispectrum to see if the parameter $g_{NL}$ could contribute to distinguishing between different models.  It is also quite clear that the machinery used in this paper leads to rather a lot of long equations.  It would be nice to come up with a way of simplifying the calculations such that one could make more general statements about N curvatons.

\section*{Acknowledgments}
MF is very grateful for funding from the STFC.  We are also very happy to thank Karim Malik, Filippo Vernizzi and David Wands.

\appendix
 \renewcommand{\theequation}{A-\arabic{equation}}
  % redefine the command that creates the equation no.
  \setcounter{equation}{0}  % reset counter
\section{Appendix}
\subsection{Case 1}

Expanding equation (\ref{mea2}) up to second order yields:
\beq
1 &=& \Omega_{\gamma_2\beta}[1 + 4({\zeta_{\gamma_2}}_{(1)} + \frac{1}{2}{\zeta_{\gamma_2}}_{(2)} - {\zeta_\beta}_{(1)} - \frac{1}{2}{\zeta_\beta}_{(2)})\nonumber\\
&& + \frac{16}{2}({\zeta_{\gamma_2}}_{(1)} - {\zeta_\beta}_{(2)})^2].
\eeq
Zeroth order is just
\be
\Omega_{\gamma_2\beta} = 1
\ee
which is an identity, as after the decay of the second curvaton all three particles have decayed into radiation.
The first order terms give simply
\be
{\zeta_\beta}_{(1)} = {\zeta_{\gamma_2}}_{(1)} ,
\ee
while the second order terms give the equation
\be
{\zeta_{\gamma_2}}_{(2)} = -2({\zeta_{\gamma_2}}_{(1)} - {\zeta_\beta}_{(2)})^2 + {\zeta_\beta}_{(2)} .
\ee
Combining these two equations we get
\be
{\zeta_{\gamma_2}}_{(2)}={\zeta_\beta}_{(2)},
\ee
which is consistent since all that's left after all the particles have decayed is radiation.

\subsubsection{Case 1 with simultaneous decay}

Below we list the nine curvature perturbation coefficients for Case 1 when we assume that the two curvatons decay at the same time. The coefficients are given in terms of $f_{\sigma_\alpha}$ and $f_{\chi_\beta}$.
\beq
{\rm A} &=&1\label{c11}\\
{\rm B} &=&\frac{f_{\sigma_\alpha}}{3}\\
\Gamma &=& \frac{1}{9}(3+f_{\sigma_\alpha})f_{\chi_\beta}\\
\Delta &=& \frac{1}{2}[-3f_{\chi_\beta}-f_{\sigma_\alpha}(3+2f_{\chi_\beta})]\\
{\rm E} &=&-\frac{1}{54}f_{\sigma_\alpha}[-9+6f_{\sigma_\alpha}(2+f_{\chi_\beta})+2f_{\sigma_\alpha}^2(3+f_{\chi_\beta})]\\
{\rm Z} &=&-\frac{1}{486}f_{\chi_\beta}[-81+6(2+f_{\sigma_\alpha})(3+f_{\sigma_\alpha})^2f_{\chi_\beta}+2(3+f_{\sigma_\alpha})^3f_{\chi_\beta}^2]\\
{\rm H} &=& -f_{\sigma_\alpha}\\
\Theta &=& -\frac{1}{3}(3+2f_{\sigma_\alpha})f_{\chi_\beta}\\
{\rm I} &=&-\frac{2}{81}f_{\sigma_\alpha}(3+f_{\sigma_\alpha})f_{\chi_\beta}[3(2+f_{\chi_\beta})+f_{\sigma_\alpha}(3+f_{\chi_\beta})]\label{c19}
\eeq

\subsection{Case 2}

The nine coefficients needed in the $f_{NL}$ expresion (\ref{gfnl}) are listed below:
\beq
{\rm A} &=&1\label{c21}\\
{\rm B} &=&\frac{1}{3}R_1f_{\sigma_\alpha}(1-f_{\chi_\beta})\\
\Gamma &=& \frac{1}{3}[1-R_1(1-f_{\chi_\alpha})(1-f_{\chi_\beta})]\\
\Delta &=&\frac{3}{2}[-1+R_1-R_1f_{\sigma_\alpha}(1-f_{\chi_\beta})-R_1f_{\chi_\alpha}(1-f_{\chi_\beta})-R_1f_{\chi_\beta}+R_1f_{\sigma_\alpha}(1 - \frac{cc^{\prime\prime}}{c^{\prime2}})\nonumber\\
&& + (1-R_1(1-f_{\chi_\alpha}))(1 - \frac{dd^{\prime\prime}}{d^{\prime2}})+f_{\chi_\beta}\frac{cc^{\prime\prime}}{c^{\prime2}}]\\
{\rm E} &=&\frac{1}{18}f_{\sigma_\alpha}(1-f_{\chi_\beta})[5R_1+2R_1f_{\sigma_\alpha}^2+2f_{\sigma_\alpha}(-1-4R_1^2+3R_1^2f_{\chi_\beta}+R_1^2f_{\chi_\beta}^2)]\nonumber\\
&& + \frac{1}{6}R_1f_{\sigma_\alpha}\frac{cc^{\prime\prime}}{c^{\prime2}}\\
{\rm Z} &=&\frac{1}{9}\{-\frac{3f_{\chi_\beta}}{2}+\frac{1}{2}(1-f_{\chi_\beta})[-5+13R_1-8R_1^2\nonumber\\
&& + 6R_1^2f_{\chi_\beta}+2R_1^2f_{\chi_\beta}^2+f_{\chi_\alpha}(4-17R_1+16R_1^2-12R_1^2f_{\chi_\beta}-4R_1^2f_{\chi_\beta}^2)]\nonumber\\
&& + 2f_{\chi_\alpha}^2(-1+2R_1-4R_1^2+R_1f_{\sigma_\alpha}+3R_1^2f_{\chi_\beta}+R_1^2f_{\chi_\beta}^2)]\}\nonumber\\
&&+ \frac{1}{6}[(1-R_1(1-f_{\chi_\alpha}))\frac{dd^{\prime\prime}}{d^{\prime2}}+f_{\chi_\beta}\frac{dd^{\prime\prime}}{d^{\prime2}}]\\
{\rm H} &=& - R_1f_{\sigma_\alpha}(1-f_{\chi_\beta})+R_1f_{\sigma_\alpha}\frac{cc^{\prime\prime}}{c^{\prime2}}\\
\Theta &=&-1+R_1 - R_1f_{\chi_\alpha}(1-f_{\chi_\beta})-R_1f_{\chi_\beta}+(1-R_1(1-f_{\chi_\alpha}))\frac{dd^{\prime\prime}}{d^{\prime2}}+f_{\chi_\beta}\frac{dd^{\prime\prime}}{d^{\prime2}}\\
{\rm I} &=&\frac{2}{9}f_{\sigma_\alpha}(1-f_{\chi_\beta})[1-5R_1+4R_1^2-3R_1^2f_{\chi_\beta}\nonumber\\
&&- R_1^2f_{\chi_\beta}^2 + f_{\chi_\alpha}(-1+R_1-4R_1^2+R_1f_{\sigma_\alpha}+3R_1^2f_{\chi_\beta}+R_1^2f_{\chi_\beta}^2)]\label{c29}
\eeq

\subsubsection{Case 2 with simultaneous decay}

With the assumption that the inflaton $\phi$ and the curvaton $\chi$ decay simultaneously we can simplify the coefficients ${\rm A}, {\rm B}, \Gamma,...$  and express in the following way:
\beq
{\rm A} &=& 1\label{c2sd1}\\
{\rm B} &=& \frac{f_{\sigma1}}{3}\\
\Gamma &=&\frac{1}{3}\frac{(4-f_{\sigma_\alpha})f_{\chi_\beta}}{(3+f_{\chi_\beta})}\\
\Delta &=&-\frac{3}{2}\frac{3f_{\sigma_\alpha}+4f_{\chi_\beta}}{(3+f_{\chi_\beta})}\\
{\rm E} &=&\frac{1}{9}(\frac{5f_{\sigma_\alpha}}{2}-5f_{\sigma_\alpha}^2+f_{\sigma_\alpha}^3)\\
{\rm Z} &=&-\frac{1}{18}\frac{(4-f_{\sigma_\alpha})f_{\chi_\beta}[-9 + (21 - 14f_{\sigma_\alpha}+2f_{\sigma_\alpha}^2)f_{\chi_\beta}]}{(3+f_{\chi_\beta})^2}\\
{\rm H} &=&-f_{\sigma_\alpha}\\
\Theta &=&-\frac{(4-f_{\sigma_\alpha})f_{\chi_\beta}}{(3+f_{\chi_\beta})}\\
{\rm I} &=&-\frac{2}{9}\frac{(4-f_{\sigma_\alpha})^2f_{\sigma_\alpha}f_{\chi_\beta}}{(3+f_{\chi_\beta})}\label{c2sd9}
\eeq

\subsection{Case 3}

The nine coefficients that appear in the curvature perturbation expression for this case are:
\beq
{\rm A} &=& 1\\
{\rm B} &=&\frac{1}{3}R_1f_{\sigma_1}(1-f_{\chi_\beta})\\
\Gamma &=&\frac{1}{3}[1-R_1(1-f_{\phi_\beta})(f_{\sigma_\alpha}+f_{\chi_\alpha})]\\
\Delta &=& \frac{3}{2}[-1 - 2R_1f_{\sigma_\alpha}^3(1-f_{\phi_\beta})+R_1(1-f_{\phi_\beta})f_{\chi_\alpha}+2R_1f_{\sigma_\alpha}^2(1-f_{\phi_\beta})(1-2f_{\chi_\alpha})]\\
{\rm E} &=& \frac{1}{18}f_{\sigma_\alpha}(1-f_{\phi_\beta})[5R_1-4R_1f_{\sigma_\alpha}^2+2f_{\sigma_\alpha}(-1+3R_1-4R_1^2+3R_1^2f_{\phi_2}+R_1^2f_{\phi_\beta}^2-6R_1f_{\chi_\alpha})]\\
{\rm Z} &=& \frac{1}{9}\{-\frac{3f_{\phi_\beta}}{2} + \frac{1}{2}(1-f_{\phi_2})[-3+2R_1f_{\sigma_\alpha}^3+9R_1f_{\chi_\alpha}\nonumber\\
&& +2(-1+2R_1-4R_1^2+3R_1^2f_{\phi_\beta}+R_1^2f_{\phi_\beta}^2)f_{\chi_\alpha}^2+2f_{\sigma_\alpha}^2(-1-4R_1^2+3R_1^2f_{\phi_\beta}+R_1^2f_{\phi_\beta}^2+2R_1f_{\chi_\alpha})\nonumber\\
&& + f_{\sigma_\alpha}(11R_1+4(-1+R_1-4R_1^2+3_1^2f_{\phi_\beta}+R_1^2f_{\phi_\beta}^2)f_{\chi_\alpha}+2R_1f_{\chi_\alpha}^2)]\}\\
{\rm H} &=& - R_1f_{\sigma_\alpha}(1-f_{\phi_\beta})[1+2f_{\sigma_\alpha}^2-2f_{\sigma_\alpha}(1-2f_{\chi_\alpha})]\\
\Theta &=& -1+R_1f_{\sigma_\alpha}(1-f_{\phi_\beta})+R_1f_{\chi_\alpha}(1-f_{\phi_\beta})\\
{\rm I} &=& -\frac{2}{9}f_{\sigma_\alpha}(1-f_{\phi_\beta})[4R_1+R_1f_{\sigma_\alpha}2+(-1+R_1-4R_1^2+3R_1^2f_{\phi_\beta}+R_1^2f_{\phi_\beta}^2)f_{\chi_\alpha}\nonumber\\
&&+f_{\sigma_\alpha}(-1-4R_1^2+3R_1^2f_{\phi_\beta}+R_1^2f_{\phi_\beta}^2+R_1f_{\chi_\alpha})]
\eeq

\subsection{Case 1*:  $t_{\sigma_{osc}}< t_{\phi_{decay}} < t_{\sigma_{decay}} < t_{\chi_{osc}} < t_{\chi_{decay}}$}

This is very similar to the first case we studied in the main body of our paper, except the curvaton $\chi$ now remains frozen until both $\phi$ and $\sigma$ decayed. Following the same process as before, we find the chain of four equations describing our model until the last decay.
Before $\sigma$ decays, the decay of the inflaton has created the background radiation while $\chi$ has not started oscillating yet and so its energy density hasn't changed from $\bar\rho_\chi$:
\be
\Omega_{\gamma_0\alpha}e^{4(\zeta_{\phi}-\zeta_\alpha)}+\Omega_{\sigma\alpha}e^{3(\zeta_\sigma-\zeta_\alpha)}+\Omega_{\chi\alpha}=1 .
\ee
Once $\sigma$ decays it becomes part of the radiation too:
\be
\Omega_{\gamma_1\alpha}e^{4(\zeta_{\gamma_1}-\zeta_\alpha)}+\Omega_{\chi\alpha}=1 .
\ee
When $\chi$ begins moving, but before it decays we have:
\be
\Omega_{\gamma_1\beta}e^{4(\zeta_{\gamma_1}-\zeta_\beta)}+\Omega_{\chi\beta}e^{3(\zeta_{\chi}-\zeta_\beta)}=1
\ee
which is the same as equation (\ref{meb2}), as the sytem is now the same as in case 1. Finally, after the last decay the valid equation is just (\ref{mea2}).

We expand these equations to second order to finally find the curvature perturbation at last decay:
\beq
\zeta_\beta&=& f_{\gamma_{0\alpha}}(1-f_{\chi_\beta}){\zeta_{\gamma_0}}_{(1)}+f_{\sigma_\alpha}(1-f_{\chi_\beta}){\zeta_{\sigma}}_{(1)}+f_{\chi_\beta}{\zeta_{\chi}}_{(1)}\nonumber\\
&& +4(1-f_{\chi_\beta})({\zeta_{\gamma_1}}_{(1)}-{\zeta_\beta}_{(1)})^2+3f_{\chi_\beta}({\zeta_{\chi}}_{(1)}-{\zeta_\beta}_{(1)})^2\nonumber\\
&& +(1-f_{\chi_\beta}){\zeta_{\gamma_1}}_{(2)}+f_{\chi_\beta}{\zeta_{\chi}}_{(2)} .
\eeq
Once we rewrite this in the form of (\ref{gcp}) it results in the following nine coefficients:
\beq
{\rm A} &=&1\\
{\rm B} &=&\frac{1}{3}f_{\sigma_\alpha}(1-f_{\chi_\beta})\\
\Gamma &=&\frac{f_{\chi_\beta}}{3}\\
\Delta &=& \frac{3}{2}(1-f_{\chi_\beta})[-(1-f_{\sigma_\alpha})f_{\sigma_\alpha}^2-f_{\sigma_\alpha}(1 - \frac{cc^{\prime\prime}}{c^{\prime2}})-f_{\chi_\beta}(1 - \frac{dd^{\prime\prime}}{d^{\prime2}})]\\
{\rm E} &=&\frac{1}{18}f_{\sigma_\alpha}[8+5f_{\sigma_\alpha}^2+f_{\sigma_\alpha}(-13 + 6f_{\chi_\beta}+2f_{\chi_\beta}^2)-3(1 - \frac{cc^{\prime\prime}}{c^{\prime2}})] \\
{\rm Z} &=& \frac{1}{9}f_{\chi_\beta}(3+f_{\chi_\beta})-\frac{1}{6}f_{\chi_\beta}(1 - \frac{dd^{\prime\prime}}{d^{\prime2}})\\
{\rm H} &=&-f_{\sigma_\alpha}(1 - \frac{cc^{\prime\prime}}{c^{\prime2}}) - (1-f_{\sigma_\alpha})f_{\sigma_\alpha}^2\\
\Theta &=& 0\\
{\rm I} &=& -\frac{2}{9}f_{\sigma_\alpha}f_{\chi_\beta}(3+f_{\chi_\beta})
\eeq

\subsection{Case 2*:  $t_{\sigma_{osc}} < t_{\sigma_{decay}} < t_{\phi_{decay}} < t_{\chi_{osc}} < t_{\chi_{decay}}$}

In the final scenario that we study, the curvaton $\chi$ still remains frozen until the other two particles have decayed. The inflaton is no longer the first one to decay, making this a modified version of case 2. The equation below, is valid just before $\phi$ decays, so the radiation component has already been created by the $\sigma$ decay:
\be
\Omega_{\gamma_0\alpha}e^{4(\zeta_{\phi}-\zeta_\alpha)}+\Omega_{\sigma\alpha}e^{3(\zeta_\sigma-\zeta_\alpha)}+\Omega_{\chi\alpha}=1 ,
\ee
and once the inflaton decays as well we have the equation:
\be
\Omega_{\gamma_1\alpha}e^{4(\zeta_{\gamma_1}-\zeta_\alpha)}+\Omega_{\chi\alpha}=1 .
\ee
Eventually, $\chi$ begins to oscillate and so its energy density will evolve like pressureless matter and equation (\ref{meb2}) is valid to describe the era until just before it decays. After $\chi$ decays we have equation (\ref{mea2}) since all that's left is radiation. The primordial curvature perturbation we obtain by second order expansion of these equations is:
Once we rewrite this in the form of (\ref{gcp}) it results in the following nine coefficients:
\beq
{\rm A} &=&1\\
{\rm B} &=&\frac{1}{3}f_{\sigma_\alpha}(1-f_{\chi_\beta})\\
\Gamma &=& \frac{1}{3}f_{\chi_\beta}\\
\Delta &=&\frac{3(1-f_{\chi_\beta})}{2}[1-2f_{\sigma_\alpha}-2f_{\sigma_\alpha}^2+2f_{\sigma_\alpha}^3-f_{\chi_\beta}+f_{\sigma_\alpha}(1-f_{\chi_\beta})\frac{cc^{\prime \prime}}{{c^\prime}^2}\nonumber\\
&& +f_{\chi_\beta}(1-f_{\chi_\beta})\frac{dd^{\prime \prime}}{{d^{\prime}}^2}]\\
{\rm E} &=& \frac{(1-f_{\chi_\beta})}{18}[3+2f_{\sigma_\alpha}+8f_{\sigma_\alpha}^3+2f_{\sigma_\alpha}^2(-8+3f_{\chi_\beta}+f_{\chi_\beta}^2)]+\frac{1}{6}f_{\sigma_\alpha}(1-f_{\chi_\beta})\frac{cc^{\prime \prime}}{{c^\prime}^2}\\
{\rm Z} &=& \frac{1-f_{\chi_\beta}}{9}[-\frac{3f_{\chi_\beta}}{2}+3(1-f_{\chi_\beta})f_{\chi_\beta}+4f_{\chi_\beta}^2]+\frac{1}{6}f_{\chi_\beta}(1-f_{\chi_\beta})\\
{\rm H} &=& (1-f_{\chi_\beta})[1-2f_{\sigma_\alpha}-2f_{\sigma_\alpha}^2+2f_{\sigma_\alpha}^3+f_{\sigma_\alpha}\frac{cc^{\prime \prime}}{{c^\prime}^2}]\\
\Theta &=& -f_{\chi_\beta}(1-f_{\chi_\beta})(1-\frac{dd^{\prime \prime}}{{d^\prime}^2})\\
{\rm I} &=& \frac{(1-f_{\chi_\beta})}{9}[-6f_{\sigma_\alpha}(1-f_{\chi_\beta})f_{\chi_\beta}-8f_{\sigma_\alpha}f_{\chi_\beta}^2]
\eeq

\end{document}